\documentclass[twocolumn,aps,pra,superscriptaddress,amsfonts,longbibliography]{revtex4-1}
\usepackage[colorlinks,linkcolor=blue,urlcolor=red,citecolor=red]{hyperref}
\usepackage{graphicx,amsmath,amsfonts,amssymb,capt-of}
\usepackage[tight]{subfigure}
\usepackage{siunitx}
\usepackage{bigints}
\usepackage{tabularx}
\usepackage{mathtools}
\usepackage{float}
\usepackage{multirow}
\usepackage{newfloat}
\usepackage{array}
\usepackage{cleveref}
\usepackage{tikz}
\usetikzlibrary{quantikz}
\usepackage{listofitems}
\usetikzlibrary{decorations.pathreplacing}
\usetikzlibrary{colorbrewer}
\usetikzlibrary{arrows}
\usetikzlibrary{arrows.meta}
\usetikzlibrary{calc}
\usetikzlibrary{positioning}
\usepackage{makecell}
\usepackage{comment}
\usepackage{physics}
\usepackage[utf8]{inputenc}
\usepackage[super]{nth}
\usepackage{amsmath}
\usepackage{geometry}
\usepackage{xcolor}
\definecolor{color1bg}{HTML}{8dd3c7} 
\definecolor{color2bg}{HTML}{ffffb3} 
\definecolor{color3bg}{HTML}{bebada} 
\definecolor{color4bg}{HTML}{fb8072} 
\definecolor{color5bg}{HTML}{80b1d3} 
\usepackage{bbm}
\newcolumntype{P}[1]{>{\centering\arraybackslash}p{#1}}

\usepackage{enumitem}
\usepackage{chemfig}
\usepackage[normalem]{ulem}
\usepackage{placeins}
\usepackage{changepage}
\newcolumntype{x}[1]{>{\centering\let\newline\\\arraybackslash\hspace{0pt}}p{#1}}
\DeclareFloatingEnvironment[
    fileext=loa,
    listname=List of Algorithms,
    name=ALGORITHM,
    placement=tbhp,
]{algorithm}

\makeatletter
\newcommand{\vast}{\bBigg@{4}}
\newcommand{\Vast}{\bBigg@{5}}
\newcommand{\VAst}{\bBigg@{6}}
\makeatother

\usepackage{colortbl}

\makeatletter
\newcommand*{\addFileDependency}[1]{
  \typeout{(#1)}
  \@addtofilelist{#1}
  \IfFileExists{#1}{}{\typeout{No file #1.}}
}
\makeatother

\begin{document}
\title{Analogy between Boltzmann machines and Feynman path integrals}
\author{Srinivasan S. Iyengar}
\email{Email: iyengar@indiana.edu}
\affiliation{Department of Chemistry, and the Indiana University Quantum Science and Engineering Center (IU-QSEC),
Indiana University, Bloomington, IN-47405}
\author{Sabre Kais}
\email{Email: kais@purdue.edu}
\affiliation{Department of Chemistry, Department of Physics and Purdue Quantum Science and Engineering Institute, Purdue University, West Lafayette, Indiana 47907}
\date{\today}

\begin{abstract}
We provide a detailed exposition of the connections between Boltzmann machines commonly utilized in machine learning problems and the ideas already well known in quantum statistical mechanics through Feynman's description of the same. We find that this equivalence allows the interpretation that the hidden layers in Boltzmann machines and other neural network formalisms are in fact discrete versions of path elements that are present within the Feynman path-integral formalism. Since Feynman paths are the natural and elegant depiction of interference phenomena germane to quantum mechanics, it appears that in machine learning, the goal is to find an appropriate combination of ``paths'', along with accumulated path-weights, through a network that cumulatively capture the correct $x \rightarrow y$ map for a given mathematical problem. As a direct consequence of this analysis, we are able to provide general quantum circuit models that are applicable to both Boltzmann machines and to Feynman path integral descriptions. Connections are also made to inverse quantum scattering problems which allow a robust way to define ``interpretable'' hidden layers.
\end{abstract}

\maketitle

\section{introduction}

Neural-network ansatz trainable on a quantum and/or classical device has been used with unprecedented success in procuring a reasonable approximation of a target quantum state\cite{sajjan2022quantum}.  Of particular interest and with a great success is the use of the Restricted-Boltzmann Machine (RBM)\cite{Wiebe-Quantum-ML}.  The choice of the RBM  is due to the fact it has been proven to be a universal approximator for any probability density \cite{Wiebe-Quantum-ML,Melko2019a, Roux_RBM} and has received astonishing success in simulating a wide variety of drivers in condensed-matter-physics\cite{Strong_corr_RBM}, quantum dynamics \cite{PhysRevResearch.3.023095}, quantum chemistry \cite{sajjan2021quantum, Xia_2018,frag-AIMD-multitop,frag-AIMD-multitop-2} and even in standard classification tasks \cite{Ciliberto2017}. Prior work has also established that RBM  is capable of mimicking a volume-law entangled quantum state even when sparsely parameterized \cite{PhysRevX.7.021021}. With quadratically scaling quantum circuits available\cite{sajjan2021quantum},  RBM  also shows hints of possible quantum advantage due to proven intractability of polynomially retrieving the full distribution classically \cite{Long2010}. Graph-based\cite{frag-TN-Anup} projection operators that resolve the identity\cite{frag-QC-Harry,frag-ML-Xiao} have been used to construct RBMs for correlated electronic potential energy surfaces\cite{frag-AIMD-multitop-2} and reduce the computational complexity for classical\cite{frag-ML-Xiao} and quantum\cite{frag-QC-Harry} calculations. 

Even though RBMs (and neural networks in general) have been widely used, probing into the underlying learning mechanism and the connection to Feynman Path Integral RBM is still sparsely explored\cite{Nori2022}. In this paper we explore a deep mathematical and conceptual connection between RBM, Feynman path integrals and more generally neural networks. This step is especially critical considering that, arguably, the most general, conceptually elegant, and unifying formalism of both quantum mechanics and statistical mechanics appears through Feynman's description of path integrals\cite{hibbs,feynman-stat-mech}. For a historical view of path integrals, see Ref. \onlinecite{Klauder_2003} and for applications to other areas see Ref. \onlinecite{Kleinert-feynmanPI}. Over the years, Feynman path integrals have been the workhorse for many path integral based molecular dynamics\cite{Makri-PI-Review,monte} and Monte Carlo\cite{Thirumalai-Berne-PI,Chandler-Wolynes-PI} formalisms to compute equilibrium properties in condensed phase quantum systems\cite{PICP,PICP-2}. The real time interpretation of Feynman path-integrals have been the basis for powerful numerical procedures such as centroid molecular dynamics\cite{cmd0,cmd-JangI,cmd-JangII} and ring-polymer molecular dynamics\cite{RPMD}. 

The paper is organized as follows: In Section \ref{RBM} we present a brief summary of RBMs which is followed by developing explicit connections between Feynman path integrals, RBMs and neural networks in general in Sections \ref{FeynmanAI}, and \ref{RBM-PI}. As a direct consequence of the analysis in Section \ref{RBM-PI}, we are able to provide general quantum circuit models that are applicable to both Boltzmann machines and to Feynman path integral descriptions. Based on this description, in Section \ref{klocal}, we are able to provide a discussion on $k$-local Hamiltonians which yield full Boltzmann machine (unrestricted) and finally in Section \ref{inverse-scattering} we present one interpretation of this mathematical exposition based in inverse scattering theory. Conclusions are given in Section \ref{concl}.


\section{Restricted-Boltzmann Machine}
\label{RBM}
\begin{figure}
    \centering   \includegraphics[width=0.48\textwidth]{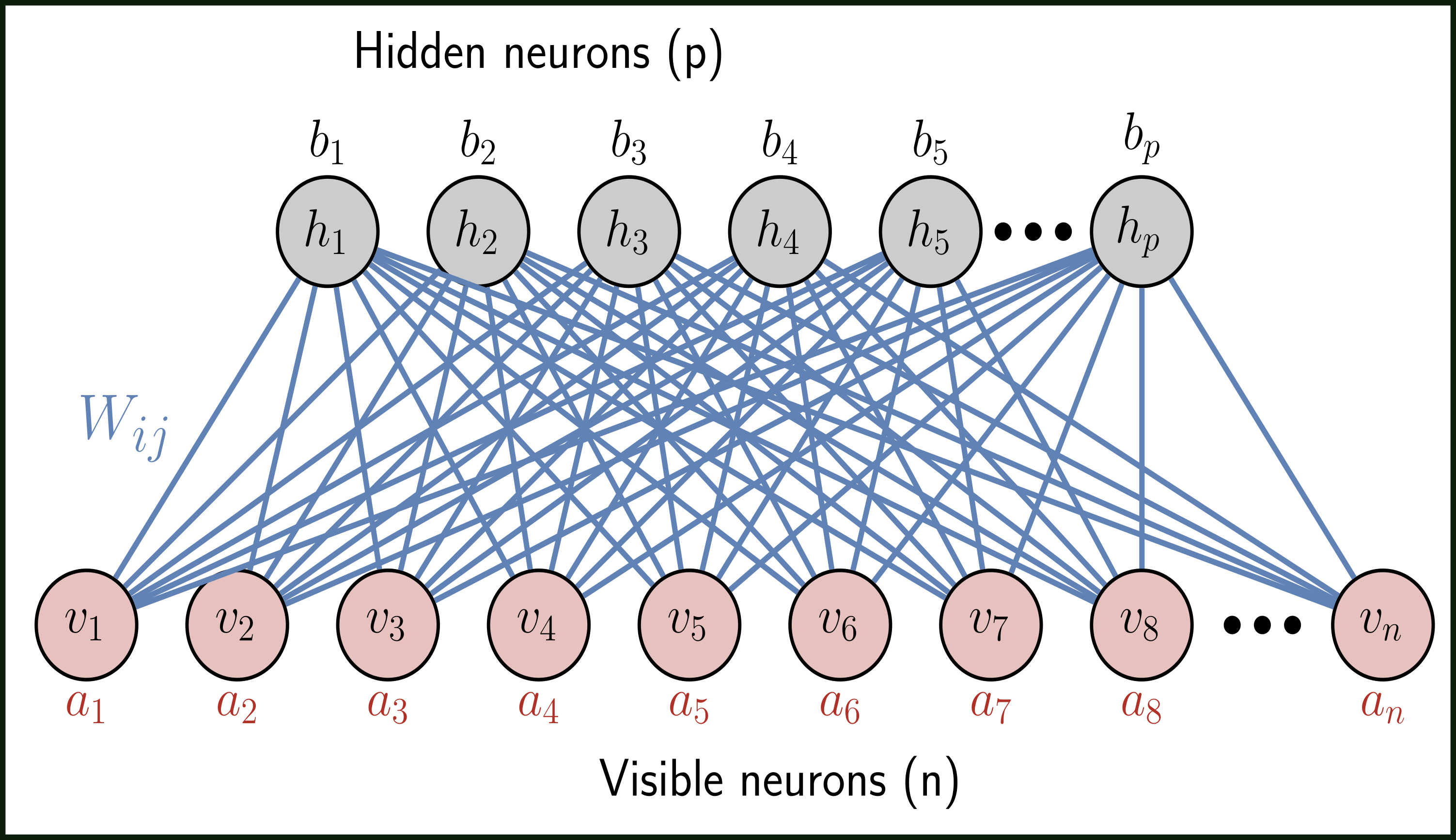}
        \caption{ RBM network G=(V,E) showing the biases $\vec{a}$ and $\vec{b}$ for hidden $\{h_j\}_{j=1}^{p}$ (grey) and visible $\{v_i\}_{i=1}^{n}$ (red) neurons and the interconnecting weights $W_{ij}$ (blue)}
    \label{RBM_scheme}
\end{figure}

The  network of the Restricted-Boltzmann Machine  denoted $G$  involves two inter-connected spin registers  $G=(V_1,V_2,E)$, where the vertex set $V_1=\{v\}_{i=1}^{i=n} \:\:\rm{with}\:\: $n$ \in \mathbb{Z}_{+}$ and each is associated with an operator $\sigma_z(v_i)$. Similar prescription exists for $V_2=\{h\}_{i=1}^{i=p} \:\:\rm{with}\:\: $p$ \in \mathbb{Z}_{+}$ and each is associated with an operator $\sigma_z(h_i)$. The network is described in the Fig. {\ref{RBM_scheme}}. The set of edges $|E|=p*n$ and is weighted by $W^i_{j}$. 
The Hamiltonian of the network is :
\begin{align}
    \mathcal{H}(\vec{X}, \vec{v}, \vec{h}) =& \sum_{i=1}^{n} a_i\sigma^z (v_i) + \sum_{j=1}^{p} b_j \sigma^z (h_j) + \nonumber \\ & \sum_{i=1,j=1}^{n,p} W^i_j\sigma^z (v_i)\sigma^z (h_j)
    \label{eq: Ising_energy}
\end{align} 
and the corresponding thermal state the network encodes is
\begin{align}
\rho^{C}_{G}(\Vec{X}, \vec{v},\vec{h}) &= \frac{e^{-\mathcal{H}(\Vec{X}, \vec{v}, \vec{h})}}
{Tr_{\{v,h\}}e^{-\mathcal{H}(\vec{X}, \vec{v}, \vec{h})}} \label{eq:rbm_dist}
\end{align} 
where the superscript $C$ denotes classically correlated thermal state. Using Eq.\ref{eq:rbm_dist} one can define a proxy
state/ansatz for the target quantum state as 
\begin{align}
\psi(\vec{X})_{H} = \sqrt{\sum_{\vec{h}} \text{diag}(\rho^{C}_{G}(\vec{X}, \vec{v}, \vec{h}))} 
\end{align} 
of the driver Hamiltonian $H \in \mathcal{C}^{d \times d}$ where $\vec{X}$ can be variationally trained.   Using the RBM network we have shown that one can obtain very accurate electronic structure of simple molecules and band structure of two-dimensional materials\cite{sajjan2021quantum, Xia_2018}. 

\section{Connections between Restricted Boltzman machines and Feynman path integrals}
\label{FeynmanAI}
The starting point in our discussion of Feynman path integrals for quantum and statistical mechanics is the quantum propagator in real and imaginary time given by 
\begin{align}
    \rho_{x,x^\prime} \equiv  \bra{x}\exp{-\beta H}\ket{x^\prime}
    \label{rhoxx'}
\end{align}
 where when $\beta$ is real,  $\rho_{x,x^\prime}$ represents the quantum canonical density operator, and $\beta$ becomes the inverse temperature or $\beta = 1/k_B T$. When $\beta$ is imaginary, that is, $\beta = \imath t /\hbar$, $\rho_{x,x^\prime}$  represents the real time evolution or unitary propagation of the problem. This process of connecting real time and imaginary inverse temperature values is commonly known as Wick rotation\cite{feynman-stat-mech} on the complex time frame, and is the hallmark of quantum statistical mechanics within Feynman's description. At this stage it is also critical to note that $\ket{x}$ may represent any basis, continuous or discreet, and we make no distinction of this in our discussion.
 
 The next step in Feynman's exposition of path integrals is to slice the propagator in Eq. (\ref{rhoxx'}) into small increments\cite{Chandler-Wolynes-PI} given by, $\delta \beta = \beta/P$. Thus for Hamiltonians that only contain two-body terms,
 \begin{align}
    \rho_{x,x^\prime} &\equiv  \bra{x} {\left[ \exp{-\delta \beta H} \right]}^P \ket{x^\prime} \nonumber \\ 
    &= \bra{x} {\left[ \exp{-\delta \beta H} \right]} {\left[ \exp{-\delta \beta H} \right]}  \cdots P \; {\text {terms}} \ket{x^\prime}
\end{align}
 a family of resolutions of identity, $\int dh_i \ket{h_i} \bra{h_i}$, inserted between the $P$ propagation slices yields
\begin{align}
    \rho_{x,x^\prime} =& 
   \int dh_1 dh_2 \cdots \bra{x}  
    {\left[ \exp{-\delta \beta H} \right]} \ket{h_1} \nonumber \\ & \phantom{\int dh_1 dh_2 \cdots} \bra{h_1} {\left[ \exp{-\delta \beta H} \right]} \ket{h_2}
     \bra{h_2} \cdots \ket{h_P} \nonumber \\ & \phantom{\int dh_1 dh_2 \cdots} \bra{h_P} {\left[ \exp{-\delta \beta H} \right]} \ket{x^\prime}  
    \label{PI-1}
\end{align}
 which has the beautiful interpretation according to Feynman\cite{feynman-stat-mech} of the particle ``traveling'' from $x$ to $x^\prime$ in a series of steps $h_1, h_2, \cdots, h_P$, which define a path, with the total amplitude on the left side being a sum over all such paths. In the discrete representation, Eq. (\ref{PI-1}) may be written as
 \begin{align}
    \rho_{x_\alpha,x_{\alpha^\prime}} =& \sum_{\bar{\mu}}
    \bra{x_\alpha}  
    {\left[ \exp{-\delta \beta H} \right]} \ket{h_1^{\mu_1}} \nonumber \\ & \phantom{\sum} \prod_{i=1}^P\bra{h_i^{\mu_i}} {\left[ \exp{-\delta \beta H} \right]} \ket{h_{i+1}^{\mu_{i+1}}}
      \nonumber \\ & \phantom{\sum} \bra{h_P^{\mu_P}} {\left[ \exp{-\delta \beta H} \right]} \ket{x_{\alpha^\prime}}  
    \label{PI-1-discrete}
\end{align}
where $\bar{\mu} \equiv \left\{ \mu_1, \mu_2, \cdots, \mu_P \right\}$ and $\mu_i$ represents the $\mu_i$-th discretization of the $i$-th slice in the Feynman path integral. That is each of the $P$ slices are discretized as noted and these discreizations are labelled using $\left\{ \mu_i \right\}$. Thus, in essence Eq. (\ref{PI-1-discrete}) is simply a discrete sum over paths labeled by  $\bar{\mu}$, or more specifically the sequence of indices, $\left\{x_\alpha,h_1^{\mu_1},h_2^{\mu_2}, \cdots h_P^{\mu_P},x_{\alpha^\prime} \right\}$ represents one specific path that connects $x$ to $x^\prime$. 
 As the number of slices $P\rightarrow \infty$, this then leads to the sum over paths notation of Feynman given by
\begin{align}
    \rho_{x,x^\prime}  
    =& \int {\cal D}h \bra{x} {\left[ \exp{-\beta H} \right]} \ket{x^\prime}
    \label{PI}
\end{align}
and the expression above is essentially a sum over paths, or path-integral over the path variable $h$, as in Eq. (\ref{PI-1-discrete}), and the path integral description is over basis vectors $\left\{ \ket{h_i} \right\}$ beginning from $\ket{x}$ and ending at $\ket{x^\prime}$; the paths are traversed by the evolution process described by the operator $\exp{-\beta H}$. Thus the interference of, or sum over,  paths, leading to the superposition theorem which is a hallmark of quantum theory, appears in Feynman's description through the accumulation of all possible paths described in Eqs. (\ref{PI-1-discrete}) and (\ref{PI}).  Additionally when the outer indices $\ket{x}$ and $\ket{x^\prime}$ are on different spaces, the expression above presents a more general path integration form for $\ket{x} \rightarrow \ket{y}$. Equations (\ref{PI-1-discrete}) and (\ref{PI}) may be compactly represented using Figure \ref{Fig:Neural-TN}. The case where $\ket{x} \rightarrow \ket{y}$ is shown in Figure \ref{Fig:Neural-TN-xy} and in a more verbose manner in Figure \ref{Fig:Neural-TN-xy-2} where the discrete version in Eq. (\ref{PI-1-discrete}) is spelt out. In all cases $\ket{x}$ and $\ket{h_i}$ represent vector spaces and hence the similarity between Figures \ref{Fig:Neural-TN}, \ref{Fig:Neural-TN-xy} and \ref{Fig:Neural-TN-xy-2}, and restricted Boltzmann machines (Figure \ref{RBM_scheme}) from machine learning is palpable. These connections will be further explored in the following sections.

\definecolor{bblue}{rgb}{0.19, 0.55, 0.91}
\definecolor{green}{rgb}{0.0, 0.65, 0.58}
\definecolor{purple}{rgb}{0.6, 0.4, 0.8}
\definecolor{orange}{rgb}{0.83, 0.4, 0.32}

\begin{figure}[h!]
    \centering

    \tikzstyle{ketx} = [circle, very thick, minimum size=1cm, inner sep=0.2mm, draw=purple, fill=purple!20]
    
    \tikzstyle{keth} = [circle, very thick, minimum size=1cm, inner sep=0.2mm, draw=orange, fill=orange!20]

    \tikzset{arw/.style={-Stealth, thick}}

    \begin{tikzpicture}[>=latex,node distance=3cm,
        every edge quote/.style={font=\fontsize{7}{1}},
        ]
		\node (x) [ketx,font=\fontsize{8}{1}]{$\ket{x}$};
        \node (h1) [keth, right=9mm of x,font=\fontsize{8}{1}]{$\ket{h_1}$};
        \node (hdots) [right=6mm of h1, minimum size=8mm,font=\fontsize{18}{1},purple]{$\cdots$};
        \node (hp) [keth, right=6mm of hdots,font=\fontsize{8}{1}]{$\ket{h_P}$};
        \node (ppx1) [below=2mm of x, minimum size=0mm, inner sep=0mm]{};
        \node (ppxN_1) [below=2mm of hp, minimum size=0mm, inner sep=0mm]{};

        \path[-]
            
           
            
            (x) edge[very thick, purple] node[above,black] {} (h1)
            (h1) edge[very thick, purple] node[above,black] {} (hdots)
            (hdots) edge[very thick, purple] node[above,black,xshift=-1mm] {}  (hp)
            (x) edge[very thick, purple] node[left,black,yshift=-1mm] {} (ppx1)
            (hp) edge[very thick, purple] node[left,black,yshift=-1mm] {} (ppxN_1)
            (ppx1) edge[very thick, purple] node[left,black,yshift=-1mm] {} (ppxN_1)
            ;

    \end{tikzpicture}

    \caption{Neural network depiction of Eq. (\ref{PI}).}
    \label{Fig:Neural-TN}
\end{figure}
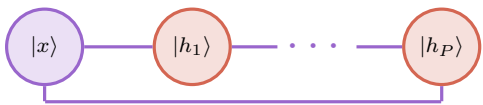
\definecolor{bblue}{rgb}{0.19, 0.55, 0.91}
\definecolor{green}{rgb}{0.0, 0.65, 0.58}
\definecolor{purple}{rgb}{0.6, 0.4, 0.8}
\definecolor{orange}{rgb}{0.83, 0.4, 0.32}

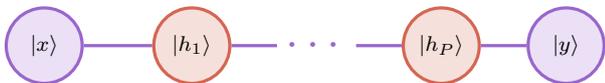
\begin{figure}[h!]
    \centering

    \tikzstyle{phi} = [circle, very thick, minimum size=1cm, inner sep=0.2mm, draw=purple, fill=purple!20]
    
    \tikzstyle{phiO} = [circle, very thick, minimum size=1cm, inner sep=0.2mm, draw=orange, fill=orange!20]

    \tikzset{arw/.style={-Stealth, thick}}

    \begin{tikzpicture}[>=latex,node distance=3cm,
        every edge quote/.style={font=\fontsize{7}{1}},
        ]
		\node (pphi1) [phi,font=\fontsize{8}{1}]{$\ket{x}$};
        \node (pphi2) [phiO, right=9mm of pphi1,font=\fontsize{8}{1}]{$\ket{h_1}$};
        \node (pphidots) [right=6mm of pphi2, minimum size=8mm,font=\fontsize{18}{1},purple]{$\cdots$};
        \node (pphiN_1) [phiO, right=6mm of pphidots,font=\fontsize{8}{1}]{$\ket{h_P}$};
        \node (pphiy) [phi, right=6mm of pphiN_1,font=\fontsize{8}{1}]{$\ket{y}$};

        \path[-]
            
           
            
            (pphi1) edge[very thick, purple] node[above,black] {} (pphi2)
            (pphi2) edge[very thick, purple] node[above,black] {} (pphidots)
            (pphidots) edge[very thick, purple] node[above,black,xshift=-1mm] {}  (pphiN_1)
            (pphiN_1) edge[very thick, purple] node[above,black] {} (pphiy)
            ;

    \end{tikzpicture}

    \caption{Neural network depiction of Eq. (\ref{PI}). Similar to Figure \ref{Fig:Neural-TN}, but now $\ket{x^\prime}$ is assumed to be a different space from $\ket{x}$. This figure is elaborated in Figure \ref{Fig:Neural-TN-xy-2} to make connections to the sum of path, Eq. (\ref{PI-1-discrete}).  }
    \label{Fig:Neural-TN-xy}
\end{figure}

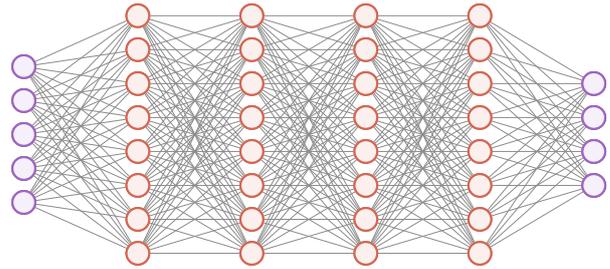
\begin{figure}[h!]
\definecolor{purple}{rgb}{0.6, 0.4, 0.8}
\definecolor{orange}{rgb}{0.83, 0.4, 0.32}
\definecolor{ggray}{rgb}{0.55, 0.55, 0.55}

\tikzstyle{mynodeV}=[thick,draw=purple,fill=purple!10,circle,minimum size=0.3cm,inner sep=0pt]
\tikzstyle{mynodeH}=[thick,draw=orange,fill=orange!10,circle,minimum size=0.3cm,inner sep=0pt]

\begin{tikzpicture}[x=1cm,y=0.75cm,scale=3]
  \readlist\Nnod{5,8,8,8,8,4} 
  \foreachitem \N \in \Nnod{ 
    \foreach \i [evaluate={\x=\Ncnt/2; \y=\N/10-0.2*\i+0.1; \prev=int(\Ncnt-1);}] in {1,...,\N}{ 
      \node[mynodeH] (N\Ncnt-\i) at (\x,\y) {};
      \ifnum\Ncnt>1 
        \foreach \j in {1,...,\Nnod[\prev]}{ 
          \draw[line width=0mm, ggray] (N\prev-\j) -- (N\Ncnt-\i); 
        }
      \fi 
    }
  }


    \readlist\Nnod{5,8,8,8,8,4} 
    \foreachitem \N \in \Nnod{ 
    \foreach \i [evaluate={\x=\Ncnt/2; \y=\N/10-0.2*\i+0.1; \prev=int(\Ncnt-1);}] in {1,...,\N}{ 
      \node[mynodeH] (N\Ncnt-\i) at (\x,\y) {};
    }
  }
  
  \readlist\Nnod{5} 
  \foreachitem \N \in \Nnod{ 
    \foreach \i [evaluate={\x=\Ncnt/2; \y=\N/10-0.2*\i+0.1; \prev=int(\Ncnt-1);}] in {1,...,\N}{ 
      \node[mynodeV] (N\Ncnt-\i) at (\x,\y) {};
      \ifnum\Ncnt>1 
        \foreach \j in {1,...,\Nnod[\prev]}{ 
          \draw[very thin] (N\prev-\j) -- (N\Ncnt-\i); 
        }
      \fi 
    }
  }
  
  \readlist\Nnod{4} 
  \foreachitem \N \in \Nnod{ 
    \foreach \i [evaluate={\x=6/2; \y=\N/10-0.2*\i+0.1; \prev=int(\Ncnt-1);}] in {1,...,\N}{ 
      \node[mynodeV] (N\Ncnt-\i) at (\x,\y) {};
    }
  }

\end{tikzpicture}
  
  \caption{\label{Fig:Neural-TN-xy-2} Explicit version of Figure \ref{Fig:Neural-TN-xy} that connects directly to sum over paths description on Eq. (\ref{PI-1-discrete}).}
\end{figure}

Furthermore, in Feynman's description, when $\beta$ is chosen to be real, that is inverse temperature ($= 1/k_B T$), the quantity,
\begin{align}
    {\cal Z} = {\text Tr} \left[ \rho \right] =
    \int dx {\cal D}h \bra{x} {\left[ \exp{-\beta H} \right]} \ket{x}
    \label{partfn}
\end{align}
is the partition function and the quantity:
$\rho_{x,x^\prime}/Z$ is canonical weight which depends on all the layers as per Eq. (\ref{PI-1}) and all possible path as per Eq. (\ref{PI-1-discrete}).

\section{Restricted Boltzmann machines recast using Feynman's path integrals}
\label{RBM-PI}
To connect the formalism above to Boltzmann machines and neural networks in general, we may begin by interpreting the family of basis states $\left\{ \ket{x} \right\}$ as {\em visible} layer states, and the layers $\left\{ \ket{h_i} \right\}$ as {\em hidden} layer states. For more general neural networks $\ket{x^\prime}$ may be replaced by $\ket{y}$. When this is not the case, there are multiple hidden layers and one visible layer, as denoted in Figures \ref{RBM_scheme} and \ref{Fig:Neural-TN}. One must note that in the traditional description of Feynman path integrals, there is no distinction between the variables used to describe the states $\left\{ \ket{x} \right\}$ and $\left\{ \ket{h_i} \right\}$ as these are both treated as belonging to the same Hilbert space. Here we choose to highlight the difference between path states, $\left\{ h_i \right\}$, and terminal states, $\left\{ \ket{x} \right\}$ to make the connections to Boltzmann machines explicit. 
In such a situation, the formalism in Eqs. (\ref{PI-1}), (\ref{PI-1-discrete}) and (\ref{PI}) yields the realization of a sum over all paths that begin at the visible basis state $\ket{x}$ and terminate at the visible basis state $\ket{x^\prime}$ by traversing through all the {\em hidden} basis points (Eq. (\ref{PI-1-discrete})) depicted as $\left\{ \ket{h_i} \right\}$. Furthermore, the evolution along these paths is dictated by $\exp{-\beta H}$, and specifically by the Hamiltonian $H$.  

To parameterize such an evolution process we may introduce the needed Hamiltonian that depicts the dynamics in Eq. (\ref{PI}) as
\begin{align}
    H =& \int dx dx^\prime \ket{x}\bra{x} H \ket{x^\prime} \bra{x^\prime} + \nonumber \\ & \sum_{i=1}^P \int dh_i \ket{h_i}\bra{h_i} H \ket{h_{i}} \bra{h_{i}} + \nonumber \\ & \int dx dh_1 \ket{x}\bra{x} H \ket{h_1} \bra{h_1} + \nonumber \\ & \sum_{i>1}^P \int dh_i dh_{i+1} \ket{h_i}\bra{h_i} H \ket{h_{i+1}} \bra{h_{i+1}} + \nonumber \\ & \int dx dh_P \ket{h_P} \bra{h_P} H \ket{x}\bra{x} + c.c.
    \label{Ham}
\end{align}
and in this general form, the connections are apparent to Eq. (\ref{eq: Ising_energy}). We note that this also represents a {\em continuous} neural network with $P$ hidden layers (compare Figures \ref{RBM_scheme} and \ref{Fig:Neural-TN}), where the diagonal elements of $H$ are biases applied to each (visible and hidden) state and the off-diagonal elements in $H$ are coupling elements across basis states, referred to as weights, either within a given visible  layer, or across neighboring layers. The discrete form of this Hamiltonian may be simply obtained by using a finite number of basis functions for  visible and hidden layers (Eq. (\ref{PI-1-discrete})) and the integrals then become summations leading to matrix elements:
\begin{align}
    {\text {Biases}}:& \left\{ \bra{x_{\alpha}} H \ket{x_{\alpha}}; \bra{h_i^{\mu_i}}  H  \ket{h_{i}^{\mu_{i}}} \right\} \nonumber \\
    {\text {Weights}}:& \left\{ \bra{x_{\alpha}} H \ket{h_{1}^{\mu_{1}}}; \bra{h_i^{\mu_i}}  H  \ket{h_{i+1}^{\mu_{i+1}}} \right\}
    \label{Ham-bias-wts}
\end{align}
and
\begin{align}
    H =
    & \sum_\alpha \bra{x_{\alpha}} H \ket{x_{\alpha}} \left\{ \ket{x_{\alpha}}\bra{x_{\alpha}} \right\} + \nonumber \\ 
    & \sum_{i,\mu_i} \bra{h_i^{\mu_i}}  H  \ket{h_{i}^{\mu_{i}}}  \left\{\ket{h_{i}^{\mu_{i}}}\bra{h_{i}^{\mu_{i}}} \right\} + \nonumber \\ 
    & \sum_{\alpha,\mu_1} \bra{x_{\alpha}}  H  \ket{h_{1}^{\mu_{1}}}  \left\{ \ket{x_{\alpha}}\bra{h_{1}^{\mu_{1}}} \right\} + c.c. + \nonumber \\ 
    & \sum_{i>1}^P \sum_{i,\mu_i,\mu_{i+1}} \bra{h_i^{\mu_i}} H \ket{h_{i+1}^{\mu_{i+1}}} \left\{ \ket{h_i^{\mu_i}} \bra{h_{i+1}^{\mu_{i+1}}} \right\} + c.c. + \nonumber \\ 
     & \sum_{\alpha^\prime,\mu_P} \bra{x_{\alpha^\prime}}  H  \ket{h_{P}^{\mu_{P}}}  \left\{ \ket{x_{\alpha^\prime}}\bra{h_{P}^{\mu_{P}}} \right\} + c.c. 
    \label{Ham-discrete}
\end{align}
which is a generalization of Eq. (\ref{eq: Ising_energy}), derived from Eq. (\ref{Ham}). In Eq. (\ref{Ham-discrete}), the terms, $\left\{ \ket{x_{\alpha}}\bra{x_{\alpha}} \right\}; \left\{\ket{h_{i}^{\mu_{i}}}\bra{h_{i}^{\mu_{i}}} \right\}; \cdots$, are projectors that are replaced by the Pauli operators in Eq. (\ref{eq: Ising_energy}). 
In such a situation, Eq. (\ref{PI}) represents the evolution process of quantum mechanics but also the learning process of machine learning. The weights for such a learning process arise from the Hamiltonian, Eqs. (\ref{Ham}), (\ref{Ham-bias-wts}) and (\ref{Ham-discrete}). Equation (\ref{Ham}) is clearly a generalization to Eq. (\ref{eq: Ising_energy}) for an arbitrary number of layers and essentially a continuous set of vertices in each layer, with Eq. (\ref{Ham-discrete}) representing the discretized version.

Additionally, we note the close connection between Eqs. (\ref{eq: Ising_energy}) and (\ref{Ham-discrete}), and the quantum Ising Model that has been studied widely on diverse quantum hardware platforms such as trapped ions \cite{monroe2021programmable}, Rydberg atoms \cite{labuhn2016tunable}, polar molecules \cite{yan2013observation}, cold atomic gases \cite{bloch2012quantum}, and superconducting circuits \cite{barends2016digitized}. In its full implementation, the quantum Ising Model Hamiltonian with local magnetic fields may be written:
\begin{equation}
H_{IT}=\sum_{\gamma}\sum_{i<j} J_{ij}^\gamma \sigma_i^\gamma \sigma_j^\gamma + \sum_{\gamma}\sum_{i}B_i^\gamma\sigma_i^\gamma
\label{IT-Hamil}
\end{equation}
where $\gamma \in {(x,y,z)}$, $J_{ij}^\gamma$ is the coupling between sites $i$ and $j$ along the $\gamma$ direction, $B_i^\gamma$ is the local magnetic field at site $i$ along the the $\gamma$ direction, and the quantities $\left\{ \sigma_i^\gamma \right\}$ are the Pauli spin operators acting on the $i^{th}$ lattice site along the $\gamma$-direction of the Bloch sphere. The critical distinction between Eqs. (\ref{eq: Ising_energy}) and (\ref{IT-Hamil}) is that, in principle all sites are connected to each other in Eq. (\ref{IT-Hamil}), and hence, $H_{IT}$ is closer to a full Boltzmann machine. 

The cost function of RBMs (see Eq. (\ref{eq:rbm_dist})) arise in Eq. (\ref{PI-1-discrete}) when $\beta$ is real. Such a situation may also be realized upon inspection of Figure \ref{Fig:Neural-TN-xy}, where each node is to be interpreted as a single layer of nodes, and the wires connecting nodes depict {\em all} weights across layers, or a linear map between the same. See Figure \ref{Fig:Neural-TN-xy-2}. These figures now summarize the analogy between Feynman's description of quantum and statistical mechanics and machine learning models as presented using Boltzmann machines. 

In the current form the Hamiltonian in Eq. (\ref{Ham}) also appears to have similarities to Ising model Hamiltonians. Compare Eqs. (\ref{Ham}), (\ref{Ham-bias-wts}), (\ref{Ham-discrete}), (\ref{IT-Hamil}) and (\ref{eq: Ising_energy}) But what is missing here is what is known as activation functions common in machine learning, which we may simply interpret as connections of the hidden bases to bath vectors, or dissipative variables, but that aspect will not be the subject of the treatment here.

\begin{figure*}[tbp]
\definecolor{purple}{rgb}{0.6, 0.4, 0.8}
\definecolor{orange}{rgb}{0.83, 0.4, 0.22}
\definecolor{ggray}{rgb}{0.55, 0.55, 0.55}

\tikzstyle{mynodeV}=[thick,draw=purple,fill=purple!10,circle,minimum size=0.2cm,inner sep=0pt]
\tikzstyle{mynodeH}=[thick,draw=orange,fill=orange!10,circle,minimum size=0.2cm,inner sep=0pt]

\subfigure[]{
\begin{tikzpicture}[x=1cm,y=0.75cm,scale=3]

\node (x1) [mynodeV,font=\fontsize{8}{1}]{$\ket{x_1}$};
\node (x1t) [right=9mm of x1]{};
\node (h11) [mynodeH, above=3mm of x1t]{$\ket{h_1^1}$}; 
\node (h12) [mynodeH, below=3mm of x1t]{$\ket{h_1^2}$}; 
\node (x2t) [right=12mm of x1t]{};
\node (h21) [mynodeH, above=3mm of x2t]{$\ket{h_2^1}$}; 
\node (h22) [mynodeH, below=3mm of x2t]{$\ket{h_2^2}$}; 
\node (x1p) [mynodeV,right=9mm of x2t,font=\fontsize{8}{1}]{$\ket{x_1^\prime}$};

\path[-]
(x1) edge[thin, purple] node[above,black] {} (h11)
(x1) edge[thin, purple] node[above,black] {} (h12)
(h11) edge[thin, purple] node[above,black] {} (h21)
(h11) edge[thin, purple] node[above,black] {} (h22)
(h12) edge[thin, purple] node[above,black] {} (h21)
(h12) edge[thin, purple] node[above,black] {} (h22)
(h21) edge[thin, purple] node[above,black] {} (x1p)
(h22) edge[thin, purple] node[above,black] {} (x1p);
\end{tikzpicture}
  }

\subfigure[]
{
\begin{quantikz}[]
   \lstick{$\ket{x_1}$} & \gate{R_y\left(\frac{\bra{x_{1}} H \ket{x_{1}}t}{\hbar}\right)} &  \ctrl{1} & \ctrl{2} & \ctrl{3} & \ctrl{4} & \qw & \meter{} \\
   \lstick{$\ket{h_1^1}$} & \gate{R_y\left(\frac{\bra{h_1^{1}}  H  \ket{h_{1}^{1}}t}{\hbar}\right)} & \ctrl{4} & \qw & \qw & \qw & \qw & \meter{} \\
   \lstick{$\ket{h_1^2}$} & \gate{R_y\left(\frac{\bra{h_1^{2}}  H  \ket{h_{1}^{2}}t}{\hbar}\right)} & \qw & \ctrl{4} & \qw & \qw & \qw & \meter{}\\
   \lstick{$\ket{h_2^1}$} & \gate{R_y\left(\frac{\bra{h_2^{1}}  H  \ket{h_{2}^{1}}t}{\hbar}\right)} & \qw & \qw & \ctrl{4} & \qw & \qw & \meter{}\\
   \lstick{$\ket{h_2^2}$} & \gate{R_y\left(\frac{\bra{h_2^{2}}  H  \ket{h_{2}^{2}}t}{\hbar}\right)} & \qw & \qw & \qw & \ctrl{4} & \qw & \meter{} \\
   \lstick{$\ket{a_{1}}$} & \qw &  \gate{R_y\left(\frac{\bra{x_1}  H  \ket{h_{1}^{1}}t}{\hbar}\right)} & \qw & \qw & \qw & \qw & \meter{} \\
   \lstick{$\ket{a_{2}}$} & \qw & \qw & \gate{R_y\left(\frac{\bra{x_1}  H  \ket{h_{1}^{2}}t}{\hbar}\right)} & \qw & \qw & \qw & \meter{} \\
   \lstick{$\ket{a_{3}}$} & \qw & \qw & \qw & \gate{R_y\left(\frac{\bra{x_1}  H  \ket{h_{2}^{1}}t}{\hbar}\right)} & \qw & \qw & \meter{} \\
   \lstick{$\ket{a_{4}}$} & \qw & \qw & \qw & \qw & \gate{R_y\left(\frac{\bra{x_1}  H  \ket{h_{2}^{2}}t}{\hbar}\right)}& \qw & \meter{} \\
   \end{quantikz}
}
  \caption{\label{Fig:Neural-TN-ckt} Illustration of the approach in Section \ref{RBM-PI} for two hidden layers and one input layer RBM system. The left visible layer is simply reproduced for convenience on the right side of Figure (a). Figure (b) not only provides a circuit model for the neural in Figure (a), but it also provides a circuit model for a Feynman path integral problem referred to in Figure (a). The $\ket{a_i}$ represent the ancilla, $\ket{x_i}$ represent the visible layer (or end points of the Feynman path) and $\ket{h_i}$ are the hidden layers (or path elements that are used to construct the superposition in the Feynman path description).}
\end{figure*}
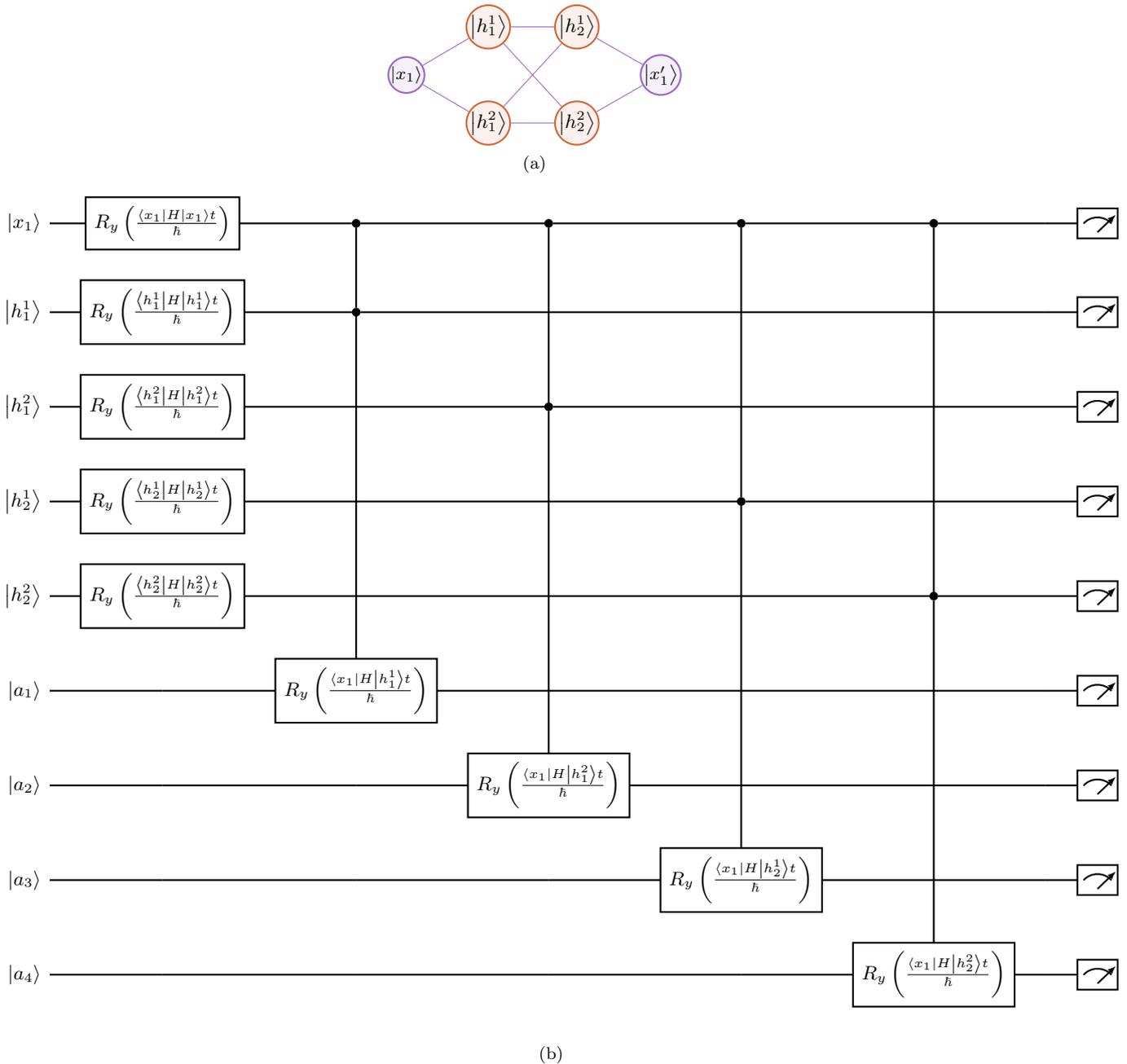

Finally, in Figure \ref{Fig:Neural-TN-ckt}, we present a circuit model for both the RBM shown and also for the Feynman path integral that is represented by Figure \ref{Fig:Neural-TN-ckt}(a). 

\subsection{Boltzmann machines as an inverse scattering problem: ``interpretability'' of the ``hidden'' layer}
\label{inverse-scattering}
In the above sections we discussed  the theory of Feynman path integrals as applicable to both quantum mechanics as well as statistical mechanics. This naturally leads to the introduction of a set of intermediate states, depicted as $\left\{ \ket{h_i} \right\}$ above, that are visited during transitions both in real time (quantum mechanics) and imaginary time (statistical mechanics and thermodynamics). We then showed how the exact same structure appears in Boltzman machines and in neural networks thus leading to one obvious interpretation that perhaps Boltzman machines are a realization of Feynman path integrals. This automatically leads to definitions for entropy through partial trace over hidden layers, in Section \ref{entropy}, and similarly definitions for higher order interactions, Sections \ref{klocal}.

We now ask if may expound upon an equivalent description for the definition of weights and biases, as alluded in Eq. (\ref{Ham-bias-wts}), as these appear in machine learning. In machine learning and Boltzmann machines, a network such as that in Figure \ref{Fig:Neural-TN-xy-2}, is represented by
\begin{align}
    \ket{h_{a+1}}=f_{a+1}({\cal W}_{a,a+1} \ket{h_{a}})
    \label{NN-eqn}
\end{align}
where $f_{a+1}$ represents the activation function for the $(a+1)$-th layer and $\left\{ {\cal W}_{a,a+1}\right\}$ are the weight tensors  (including bias) connecting the $a$-th and $(a+1)$-th layers. In defining $\left\{ {\cal W}_{a,a+1}\right\}$ here, we have combined the $\left\{ b_j;  W^i_j \right\}$ terms in Eq. (\ref{eq: Ising_energy}). The functions, $f_{a+1}$, are chosen to be functions of the kets created by the action, $\left[ {\cal W}_{a,a+1} \ket{h_{a}} \right]$, and thus the weights $\left\{ {\cal W}_{a,a+1}\right\}$ are operators that act on kets. The assumption here is that $f_i$ are analytic, differentiable functions and hence a continuous representation of ReLU are acceptable forms. For the description in Section \ref{FeynmanAI}, 
\begin{align}
\bra{h_{a+1}} {\left[ \exp{-\delta \beta H} \right]} \ket{h_{a}} \equiv \bra{h_{a+1}}f_{a+1}({\cal W}_{a,a+1} \ket{h_{a}})
\label{Feyn=AI}
\end{align}
and this is the key central insight that appears from the  treatment in Section \ref{FeynmanAI}, and crystalizes the analogy presented in this paper. Thus, influenced by Eq. (\ref{PI-1}), an equivalent description for the machine learning processes described by Figures \ref{Fig:Neural-TN-xy} and \ref{Fig:Neural-TN-xy-2} may be written in the continuous limit as,
\begin{align}
    \ket{y} =& 
   \int dh_1 dh_2 \cdots   
    f_{P+1} \left( {\cal W}_{P+1,P} \ket{h_P} \right) \nonumber \\ & \phantom{\int dh_1 dh_2 \cdots} \bra{h_P} f_P \left( {\cal W}_{P,P-1} \ket{h_{P-1}} \right)
     \bra{h_{P-1}} \cdots \ket{h_2} \nonumber \\ & \phantom{\int dh_1 dh_2 \cdots} \bra{h_2} f_{2}({\cal W}_{1,2} \ket{x}) 
    \label{ML-PI-1}
\end{align}
Note the close similarity between Eqs. (\ref{PI-1}) and (\ref{ML-PI-1}) that both seem to now have a sum over paths flavor. Clearly, whereas in quantum dynamics Eq. (\ref{PI-1}) may be computed by providing a system Hamiltonian, in machine learning, the weights, $\left\{ {\cal W}_{a,a+1}\right\}$, are to be obtained based on a known set of transitions, $\ket{y} \leftarrow \ket{x}$ that are used for training the network. 
Thus in a sense the ML approach to Boltzman machines is one where the Hamiltonian is computed based on a dataset that captures the $\ket{y} \leftarrow \ket{x}$ map. 

Hence, to achieve an equivalent description for the weights, $\left\{ {\cal W}_{a,a+1}\right\}$, we may remind ourselves that in quantum dynamics and scattering theory\cite{RGNewton}, the transition amplitude: $\bra{\chi_f} \Omega \ket{\chi_i}$ under the influence of the M\"oller operator, $\Omega$, which includes time-evolution (as in Section \ref{FeynmanAI}) or associated Greens function\cite{RGNewton}, is the key aspect and connects to various observables such as state-to-state scattering probabilities, rate constants, and also vibrational properties. By comparison this statement is not dissimilar to that in Eqs. (\ref{PI-1}) and (\ref{ML-PI-1}), but in fact from this perspective, the machine learning problem is an inverse scattering problem in that, one may say, the Hamiltonian for the process, given by Eqs. (\ref{Ham}), (\ref{Ham-bias-wts}) and (\ref{Ham-discrete}), needs to be discovered. In this language, the ``hidden'' layers of machine learning are simply the steps needed to construct the interference experiments germane to quantum mechanics. Whereas in quantum mechanics these ``hidden'' steps seem to allow the system to achieve a probabilistic view, in machine learning the same probabilistic view samples a  large parameter space thus allowing for an optimal solution to a given (hard) problem.

\section{Description of entropy}
\label{entropy}
To describe the entropy of the system containing $\left\{ \ket{x}; \left\{ \ket{h_i} \right\} \right\}$, using Eq. (\ref{partfn}), we may first define the marginal probabilities using the matrix elements 
\begin{align}
\rho_{x,h_1} = \bra{x}  
    {\left[ \exp{-\delta \beta H} \right]} \ket{h_1}
\end{align}
and
\begin{align}
\rho_{h_i,h_{i\pm 1}} = \bra{h_i}  
    {\left[ \exp{-\delta \beta H} \right]} \ket{h_{i\pm 1}}
\end{align}
to redefine Eq. (\ref{PI}) as
\begin{align}
    \rho_{x,x^\prime} = \int dh_1 dh_2 \cdots \rho_{x,h_1} \rho_{h_1,h_2} \cdots 
    \label{PI2}
\end{align}
These marginals may be used to write the Shannon entropy functions: ${\cal S}[\rho_{x,h_1}]$, and ${\cal S}[\rho_{h_i,h_{i\pm1}}]$. The overall entropy of the network is obtained using the entropy of each layer in a manner similar to the inclusion exclusion principle in set theory\cite{PIE}, an appropriate generalization for which is provided by Bethe's free-energy\cite{Belief-Kikuchi} and may be written as
\begin{align}
 {\cal S}[\rho_{x}]+&\sum_i {\cal S}[\rho_{h_i}]- {\cal S}[\rho_{x,h_1}]- \sum_i {\cal S}[\rho_{h_i,h_{i+1}}]
\label{S2}
\end{align}
which is often referred to as mutual information entropy and also has applications in Belief propagation\cite{Belief-Kikuchi}. For the special case of the network in Figure \ref{RBM_scheme}, Eq. (\ref{S2}) reduces to
\begin{align}
 {\cal S}[\rho_{x}]+ {\cal S}[\rho_{h_1}]- {\cal S}[\rho_{x,h_1}]
\label{S2-Fig1}
\end{align}
which is consistent with the expression in Ref. \onlinecite{https://doi.org/10.48550/arxiv.2208.13384}. However, Eq. (\ref{S2}) provides a genralization for an arbitrary number of hidden layers for more general RBMs beyond those in Figure \ref{RBM_scheme}.

\definecolor{bblue}{rgb}{0.19, 0.55, 0.91}
\definecolor{green}{rgb}{0.0, 0.65, 0.58}
\definecolor{purple}{rgb}{0.6, 0.4, 0.8}
\definecolor{orange}{rgb}{0.83, 0.4, 0.32}

\begin{figure}[h!]
    \centering

    \tikzstyle{kets} = [circle, very thick, minimum size=1cm, inner sep=0.2mm, draw=purple, fill=purple!20]
    
    \tikzstyle{ketsh} = [circle, very thick, minimum size=1cm, inner sep=0.2mm, draw=orange, fill=orange!20]

    \tikzstyle{U} = [rectangle, thick, minimum size=0.25cm, draw=bblue, fill=bblue!20]

    \tikzset{arw/.style={-Stealth, thick}}

    \begin{tikzpicture}[>=latex,node distance=3cm,
        every edge quote/.style={font=\fontsize{7}{1}},
        ]
		\node (x) [kets,font=\fontsize{8}{1}]{$\ket{x}$};
        \node (ptemp1) [right=9mm of x, font=\fontsize{8}{1}]{};
        \node (h1) [ketsh,above=9mm of ptemp1, font=\fontsize{8}{1}]{$\ket{h_1}$};
		\node (psq1) [U,above=4.5mm of ptemp1]{};
        \node (h2) [ketsh,right=9mm of ptemp1, font=\fontsize{8}{1}]{$\ket{h_2}$};
		\node (psq2) [U,above=4.5mm of h2]{};
        \node (h3) [ketsh,right=18mm of h1, font=\fontsize{8}{1}]{$\ket{h_3}$};
      \node (hdots) [right=18mm of h2, minimum size=8mm,font=\fontsize{18}{1},orange]{$\cdots$};
        \node (hdots2) [above=9mm of hdots, minimum size=8mm,font=\fontsize{18}{1},orange]{$\cdots$};
      \node (hp) [ketsh,right=6mm of hdots, font=\fontsize{8}{1}]{$\ket{h_P}$};
      \node (pphi1down) [below=2mm of x, minimum size=0mm, inner sep=0mm]{};
      \node (pphi5down) [below=2mm of hp, minimum size=0mm, inner sep=0mm]{};

        \path[-]
            
           
            
            (x) edge[very thick, purple] node[above,black] {} (pphi1down)
            (x) edge[very thick, purple] node[above,black] {} (h1)
            (x) edge[very thick, purple] node[above,black] {} (h2)
            (h1) edge[very thick, purple] node[above,black] {} (h2)
            (h2) edge[very thick, purple] node[above,black] {} (h3)
            (h1) edge[very thick, purple] node[above,black] {} (h3)
            (hdots) edge[very thick, purple] node[above,black] {} (hp)
            (hdots) edge[very thick, purple] node[above,black] {} (h3)
            (pphi5down) edge[very thick, purple] node[above,black] {} (pphi1down)
            (hp) edge[very thick, purple] node[above,black] {} (pphi5down)
            (hp) edge[very thick, purple] node[above,black] {} (hdots2)
            ;

    \end{tikzpicture}

    \caption{3-local Neural network depiction with entropy in Eq. (\ref{S-Kekuchi}). Each triangle represents a rank-3 weight tensor and is depicted with a blue square on its interior. In general this could be a partial tree topology but is presented here in a simplified form.}
    \label{Fig:Neural-2}
\end{figure}
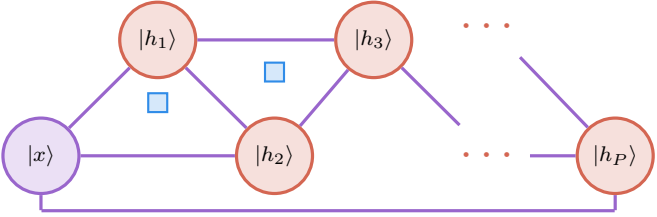
\definecolor{bblue}{rgb}{0.19, 0.55, 0.91}
\definecolor{green}{rgb}{0.0, 0.65, 0.58}
\definecolor{purple}{rgb}{0.6, 0.4, 0.8}
\definecolor{orange}{rgb}{0.83, 0.4, 0.32}

\begin{figure}[h!]
    \centering

    \tikzstyle{kets} = [circle, very thick, minimum size=1cm, inner sep=0.2mm, draw=purple, fill=purple!20]
    
    \tikzstyle{ketsh} = [circle, very thick, minimum size=1cm, inner sep=0.2mm, draw=orange, fill=orange!20]

    \tikzstyle{U} = [rectangle, thick, minimum size=0.25cm, draw=bblue, fill=bblue!20]

    \tikzset{arw/.style={-Stealth, thick}}

    \begin{tikzpicture}[>=latex,node distance=3cm,
        every edge quote/.style={font=\fontsize{7}{1}},
        ]
		\node (x) [kets,font=\fontsize{8}{1}]{$\ket{x}$};
        \node (ptemp1) [right=9mm of x, font=\fontsize{8}{1}]{};
        \node (h1) [ketsh,above=9mm of ptemp1, font=\fontsize{8}{1}]{$\ket{h_1}$};
		\node (psq1) [U,above=4.5mm of ptemp1]{};
        \node (h2) [ketsh,right=9mm of ptemp1, font=\fontsize{8}{1}]{$\ket{h_2}$};
		\node (psq2) [U,above=4.5mm of h2]{};
        \node (h3) [ketsh,right=18mm of h1, font=\fontsize{8}{1}]{$\ket{h_3}$};
      \node (hdots) [right=18mm of h2, minimum size=8mm,font=\fontsize{18}{1},orange]{$\cdots$};
        \node (hdots2) [above=9mm of hdots, minimum size=8mm,font=\fontsize{18}{1},orange]{$\cdots$};
      \node (y) [kets,right=6mm of hdots, font=\fontsize{8}{1}]{$\ket{y}$};
      \node (pphi1down) [below=2mm of x, minimum size=0mm, inner sep=0mm]{};
      \node (pphi5down) [below=2mm of y, minimum size=0mm, inner sep=0mm]{};

        \path[-]
            
           
            
            (x) edge[very thick, purple] node[above,black] {} (pphi1down)
            (x) edge[very thick, purple] node[above,black] {} (h1)
            (x) edge[very thick, purple] node[above,black] {} (h2)
            (h1) edge[very thick, purple] node[above,black] {} (h2)
            (h2) edge[very thick, purple] node[above,black] {} (h3)
            (h1) edge[very thick, purple] node[above,black] {} (h3)
            (hdots) edge[very thick, purple] node[above,black] {} (y)
            (hdots) edge[very thick, purple] node[above,black] {} (h3)
            (pphi5down) edge[very thick, purple] node[above,black] {} (pphi1down)
            (y) edge[very thick, purple] node[above,black] {} (pphi5down)
            (y) edge[very thick, purple] node[above,black] {} (hdots2)
            ;

    \end{tikzpicture}

    \caption{Same as Figure \ref{Fig:Neural-2} but now depicts the case for $\ket{x} \rightarrow \ket{y}$ as in Figure \ref{Fig:Neural-TN-xy}.}
    \label{Fig:Neural-2-xy}
\end{figure}
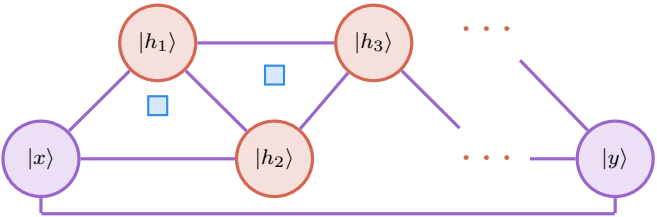
\section{${\text{k}}$-local Hamiltonians yield Boltzmann machines}
\label{klocal}
For cases where the Hamiltonian in Eq. (\ref{Ham}) has $k$-body terms, the situation in Eqs. (\ref{PI}), (\ref{PI2}) and (\ref{S2}) is more complicated. For example, for 3-body terms in Eq (\ref{Ham}), the expression in Eq. (\ref{PI2}) may be written as
\begin{align}
    \rho_{x,x^\prime} = \int dh_1 dh_2 \cdots \rho_{x,h_1,h_2} \rho_{h_1,h_2,h_3} \cdots \rho_{h_{P-1},h_P,x^\prime}
    \label{PI3}
\end{align}
The associated neural network and Feynman path integration techniques are both represented in compact form in Figure \ref{Fig:Neural-2}. Here each triangle captures to the three-body interaction and this aspect is referred to using the blue squares inside the triangles. For example, the three body terms make the Hamiltonian tensorial and thus it simultaneously interacts bases $\left\{ \vec{x}; \vec{h_1}; \vec{h_2}\right\}$, etc. While a tree-type topology may be appropriate in such cases as the interactions grow, a simplified form of the representation is presented in Figure \ref{Fig:Neural-2}.
Again, as in Figure \ref{Fig:Neural-TN-xy}, when the outer indices $\ket{x}$ and $\ket{x^\prime}$ are on different Hilbert spaces, $\ket{x^\prime}$ in Eq. (\ref{PI3}) is replaced with $\ket{y}$. The associated depiction is provided in Figure \ref{Fig:Neural-2-xy}. It must be noted that Figure \ref{Fig:Neural-2} is not  a restricted Boltzmann machine as may be seen from the fact that the layer corresponding to $\ket{x}$ is connected to two following layers corresponding to $\ket{h_1}$ and $\ket{h_2}$, and so on. 
In fact this is a step towards a general Boltzmann machine and as the many-body interactions captured within the Hamiltonian increases, this approaches the path-integral formalism commensurate with the full Boltzmann machine. 

In such cases a generalization to the entropy from Eq. (\ref{S2}) may be obtained from Kekuchi's theory\cite{Belief-Kikuchi}. We begin this generalization by reinspecting Figures \ref{Fig:Neural-2} and \ref{Fig:Neural-2-xy}. These figures contain sets of triangles that are connected to each other since the Hamiltonian contains three-body terms. Thus higher order Hamiltonians will necessitate the presence of higher order simplexes\cite{DEY1997267} that are connected, and these will be commensurate with the many-body interaction terms that are captured within the Hamiltonian. Thus, it is appropriate to think of the Boltzmann machine in  Figures \ref{Fig:Neural-2} and \ref{Fig:Neural-2-xy} as graphs such that when the Hamiltonian contains  $k$-body terms, the associated Boltzmann machine would have $k$-nodes that are completely connected and hence best represented as $k$-simplexes. 
Thus if we consider the resultant neural network as a graph made of simplexes, or as a simplicial complex\cite{DEY1997267,armstrong2013basic}, the entropy arises from a graph theoretic description\cite{Belief-Kikuchi,fragPBC,frag-TN-Anup} and may be written as
\begin{align}
\sum_{\alpha,r}^{\cal R} (-1)^r {\cal S}_{\alpha,r} {\cal M}_{\alpha,r}
\label{S-Kekuchi}
\end{align}
Here the $\alpha$-th rank-$r$ simplex within the graph created from the neural network has entropy ${\cal S}_{\alpha,r}$, which may be thought to be a functional of the reduced probability $\rho_{\alpha,r}$, that is the reduced probability for the $\alpha$-th rank-$r$ simplex in a graphical depiction such as that in Figures \ref{Fig:Neural-2} and \ref{Fig:Neural-2-xy}. The quantity ${\cal R}$ in Eq. (\ref{S-Kekuchi}) is the maximum rank of the simplexes that is the ${\cal R}=k$ for $k$-local Hamiltonians. Thus, for the two-body case, the rank of the objects within a graphical description of the neural network is ``$r=0$'' (nodes in the network) and ``$r=1$'' (edges in the network). For the three-body case in Figures \ref{Fig:Neural-2} and \ref{Fig:Neural-2-xy}, the value of ${\cal R}$ is 2 and the corresponding entropy takes the form
\begin{align}
\sum_{\alpha \in {\text {nodes}}} {\cal S}_{\alpha,0} {\cal M}_{\alpha,0} - \sum_{\alpha \in {\text {edges}}} {\cal S}_{\alpha,1} {\cal M}_{\alpha,1} + \sum_{\alpha \in {\text {faces}}} {\cal S}_{\alpha,2} {\cal M}_{\alpha,2}
\label{S-Kekuchi-2}
\end{align}
where the set of ``nodes'' include visible and hidden layers, the set of ``edges'' include connections between the same and are determined by the traditional weights used in machine learning and finally the set of ``faces'' include generalized weights that now depend on three sets oflayers that may be any combination of visible or hidden layers. 

The quantity, ${\cal M}_{\alpha,r}$ in Eq. (\ref{S-Kekuchi}) is a multiplicity term and prevents over-counting in the graph-theoretic expression, Eq. (\ref{S-Kekuchi}), and includes the number of times the $\alpha$-th  $(r)$-rank object appears in all simplexes of rank greater than or equal to $r$. Thus through the analogy discussed we are also able to provide high-order neural networks as a extension of Feynman path integrals with Hamiltonians that may contain higher order terms. 

\section{conclusion}
\label{concl}
Machine learning has had great impact recently in a number of areas of science. Recently quantum versions of machine learning protocols have also been constructed. As machine learning grows in impact, there has been a wide discussion in the literature that deal with the interpretation of hidden layers as they appear in these formalisms. In this paper, we provide a general description for many problems in machine learning, and more precisely Boltzmann machines, by finding an analogy between these and the Feynman path integral description of quantum and statistical mechanics. We find that the basic mathematical structure of RBMs reminds us of a superposition of (or sum over) paths structure, which is a critical hallmark of Feynman's description of quantum and statistical mechanics. This then allows us to reinterpret the hidden layers in machine learning as being akin to the intermediate, or virtual, states visited by quantum systems as part of the path integration for quantum propagation in real and imaginary time. As a direct consequence of this argument, we are able to introduce a general quantum circuit that encompasses both RBMs and Feynman path integrals. 

We then find that while 2-local Hamitonians within the Feynman path integral formalism are reminiscent of RBMs, $k$-local Hamiltonians naturally yield a structure that looks like a Boltzmann machines without restrictions. In fact in such cases the neural networks obtained look more like a simplical complex to allow connections that go beyond nearest neighbor of hidden layers. Given the isomorphism to graphs and simplicial complexes, we are also able to provide general expressions for entropy by applying the inclusion-exclusion principle directly to the simplicial complexes. 

We have recently demonstrated \cite{https://doi.org/10.48550/arxiv.2208.13384} that how information between the two spin-registers of the network can flow in real-time during training and how such a finding can be leveraged to identify robust yet emergent training principles. We have further analytically related such information transport quantifiers to usual measures of correlation and have established rigorous bounds satisfied by the two-quantities.  Connection to Feynamn Path Integrals might shed light on such training dynamics and open a new path to analyze  how footprints of quantum correlation within the physical system studied gets imprinted onto the learner network.

\section{Acknowledgment}

This research was supported by the National Science
Foundation grants CHE-2102610 and OMA-1936353 to S.S.I. 
S.K. would  like to thank  Dr. Manas Sajjan for many useful discussions and acknowledges the National
Science Foundation under Award No. 1955907 and the U.S. Department of Energy (Office of Basic Energy Sciences) under Award No. DESC0019215. SSI acknowledges Mr. Xiao Zhu for his useful comments on the paper.


\end{document}